# Deterministic Field-free Switching of a Perpendicularly Magnetized Ferromagnetic Layer via the Joint Effects of Dzyaloshinskii-Moriya Interaction and Field-like Spin-orbit Torque: An Appraisal

Kai Wu[1], Diqing Su[2], Renata Saha[1], and Jian-Ping Wang[1, *]

[1]Department of Electrical and Computer Engineering, University of Minnesota, Minneapolis, Minnesota 55455, USA

[2]Department of Chemical Engineering and Material Science, University of Minnesota, Minneapolis, Minnesota 55455, USA

*Corresponding author E-mail: jpwang@umn.edu

**Abstract:** Field-free switching of perpendicularly magnetized ferromagnetic layer by spin orbit torque (SOT) from the spin Hall effect (SHE) is of great interest in the applications of magnetic memory devices. In this paper, we investigate the deterministic SOT switching through the joint effects of Dzyaloshinskii-Moriya Interaction (DMI) and field-like torque (FLT) by micromagnetic simulations. We confirmed that within a certain range of DMI values and charge current densities, it is possible to deterministically switch the magnetization without the assistance external magnetic field. We show that the FLT could play an adverse role in blocking and slowing down the magnetization switching under certain cases of DMI and charge current-driven field-free switching. However, in other cases, FLT can assist DMI on the deterministic field-free SOT switching. In addition, it is found that FLT can effectively expand the current density window for a deterministic field-free SOT switching in the presence of DMI.

## 1. Introduction

Recently, there have been extensive experiments on the demonstration of deterministic switching of perpendicularly magnetized layers in heavy metal/ferromagnet (HM/FM) structures with large spin-orbit coupling, offering a new route towards high speed and low power storage applications [1-7]. Specifically, in this HM/FM structure, an in-plane charge current is applied to the HM layer to generate an out-of-plane spin polarized current due to the strong spin-orbit interactions. The spin torque from the spin polarized current is transferred to the FM layer and induces magnetization switching. This switching mechanism has led to the spin-orbit torque magnetic random access memory (MRAM). The spin torque contributes to the magnetic dynamics in the FM layer and it is described by the Landau-Lifshitz-Gilbert-Slonczewski (LLGS) equation. However, the injected spin polarized current simply drives the magnetization from out-of-plane to in-plane, so conventional spin-orbit torque (SOT) switching requires an external in-plane magnetic field that is collinear with the charge current for stable bidirectional magnetization switching, which is a major challenge for practical realization of devices [8]. Recently, the ability to perform SOT switching without an external field is of great interest for applications. To realize field-free SOT switching of a perpendicular magnetization, many works have been reported such as adding a bias layer [9-11]. However, this complicates the fabrication process and potentially reduces the memory density [12]. Other methods such as using the assistance of an antiferromagnetic layer (AFM) [13, 14], a tilted anisotropy [15], and broken lateral inversion symmetry [16] have also been reported.

On the other hand, it has been shown that the Dzyaloshinskii-Moriya Interaction (DMI) is playing an important role in SOT switching [17, 18] as well as domain wall motion [19]. The chiral effective field arises from DMI at the FM and HM interfaces, along with the spin polarized current, may contribute to the deterministic field-free switching of a perpendicularly magnetized FM layer [18, 20-22]. Furthermore, Slonczewski-like (damping-like) torque (SLT or DLT) has been mainly used to analyze the SOT switching and domain wall dynamics. However, the role of field-like torque (FLT) has not been paid much attention in most of the SOT switching works. Until now, the role of FLT in SOT switching is remain vague. Recent works have pointed out that the FLT can be much larger than SLT in some material combinations and it is possible to use FLT to switch the magnetization [23, 24]. In this regard, we investigate the role of DMI and FLT in deterministic field-free SOT switching. This paper focuses on the dynamic switching in perpendicularly magnetized multidomain FM layer (circular dot with diameter of 100 nm) via the joint effects of DMI and FLT. Our results show that for certain FLT values, it can assist in the DMI induced deterministic field-free SOT switching, for other cases, FLT impedes and slows down the deterministic switching.

## 2. Theoretical Approach

The Landau-Lifshitz-Gilbert-Slonczewski (LLGS) equation including DMI and SOT terms is expressed as:

$$\frac{\partial \boldsymbol{m}}{\partial t} = -|\gamma|\mu_0 \boldsymbol{m} \times \boldsymbol{H}_{eff} + \alpha \boldsymbol{m} \times \frac{\partial \boldsymbol{m}}{\partial t} + \tau_{SL} + \tau_{FL}$$

$$\boldsymbol{H}_{eff} = \boldsymbol{H}_K + \boldsymbol{H}_d + \boldsymbol{H}_e + \boldsymbol{H}_{DMI}$$

Where $\boldsymbol{m} = \frac{\boldsymbol{M}}{M_s}$ is the unit magnetization vector with $M_s$ the saturation magnetization, $\boldsymbol{H}_{eff}$ is an effective field including the anisotropy field $\boldsymbol{H}_K$, demagnetization field $\boldsymbol{H}_d$, exchange field $\boldsymbol{H}_e$, and the interfacial anisotropic exchange field $\boldsymbol{H}_{DMI}$ due to the Dzyaloshinskii-Moriya interaction (DMI), $\gamma$ is the gyromagnetic ratio, $\alpha$ is the Gilbert damping parameter, $\mu_0$ is the vacuum permeability constant. The SOT can be described as the sum of longitudinal Slonczewski-like (damping-like) torque $\tau_{SL}$ and the transverse field-like torque $\tau_{FL}$.

$$\tau_{SL} = |\gamma|\beta\epsilon(\boldsymbol{m} \times (\boldsymbol{\sigma} \times \boldsymbol{m}))$$

$$\tau_{FL} = |\gamma|\beta\epsilon'(\boldsymbol{\sigma} \times \boldsymbol{m})$$

$$\beta = \frac{\hbar J_e}{|e|M_s t_F}$$

Where $\hbar$ is the reduced Planck constant, $e$ is the elementary charge, $t_F$ is the thickness of FM layer, $J_e$ is the charge current density flowing in the HM (we assume $J_e$ flows along x-axis in this paper). We describe the magnitudes of $\tau_{SL}$ and $\tau_{FL}$ by their efficiencies $\epsilon$ and $\epsilon'$. In a HM/FM structure, $\epsilon = \frac{\theta_{SHA}}{2}$, where $\theta_{SHA}$ is the spin Hall angle of HM layer. The directions of SOTs are determined by the spin polarization $\boldsymbol{\sigma}$ of the injected electrons, which, from the spin Hall effect is perpendicular to the spin current and the charge current directions (in Figure 1(a), it is either along +y or -y direction).

In this work, we are exploring the joint effects of DMI and field-like torque $\tau_{FL}$ on the deterministic field-free switching of a perpendicularly magnetized ferromagnetic layer. The typical values of DMI varies from different works reported [25, 26]. Herein, we are assuming the DMI values range from 0 to 1 $mJ/m^2$. The contribution of field-like torque $\tau_{FL}$ is characterized by a dimensionless ratio $R$:

$$R = \frac{\tau_{FL}}{\tau_{SL}}$$

Herein, we perform simulation using the Object Oriented Micro-Magnetic Framework (OOMMF) public code [27, 28] by numerically solving the LLGS equation. The parameters used in this OOMMF simulation are listed in Table 1. The FM layer is assumed to be circular with a diameter of 100 nm, thickness of 1 nm, and is discretized into 2 nm × 2 nm × 1 nm for calculations. During all the simulations in this paper, the device is initially relaxed for 3 ns before a current with a pulse width of $t_p = 5\ ns$ is applied, followed by another 5 ns of relaxation. So, the magnetic dynamics in FM layer is recorded in a 13 ns time window.

Table 1. Simulation Parameters

| Parameter | Description | Values |
| --- | --- | --- |
| **FM Layer Dimensions** | Diameter × Thickness | 100 nm × 1 nm |
| **Cell size** | Length × Width × Thickness | 2 nm × 2 nm × 1 nm |
| $\gamma$ | Gyromagnetic ratio | $2.21 \times 10^5 \ m/A \cdot s$ |
| $\alpha$ | Gilbert damping factor | 0.1 |
| A | Exchange constant | $2 \times 10^{-11} \ J/m$ |
| $M_s$ | Saturation magnetization | $1.1 \times 10^6 \ A/m$ |
| $\theta_{SHA}$ | Spin Hall angle | 0.3 |
| $K_u$ | Intrinsic perpendicular anisotropy of FM layer | $8 \times 10^5 \ J/m^3$ |
| $H_K$ | Perpendicular anisotropy field | 1.45 T |
| **DMI** | Dzyaloshinskii-Moriya interaction factor | $0 - 1.0 \ mJ/m^2$ |
| $R\left(=\dfrac{\tau_{FL}}{\tau_{SL}}=\dfrac{\epsilon'}{\epsilon}\right)$ | Ratio of the field-like torque to the Slonczewski-like torque | 0, 0.01, 0.03, 0.05, 0.1, 0.3 |
| $J_e$ | Charge current density | $0.5 - 1.5 \times 10^{12} \ A/m^2$ |

It has been reported that the DMI assists in the deterministic field-free SOT switching of a perpendicular ferromagnet [21]. Figure 1(b) shows the simulated temporal evolution of the perpendicular magnetization component $m_z$ with and without the assistance of the DMI term. For the case with DMI = 0.5 $mJ/m^2$, the $m_z$ reaches to negative (-z direction) within 1 ns from the onset of charge current, resulting in a deterministic magnetization switching in FM layer. However, under the same condition but without the presence of DMI term, the $m_z$ approaches to 0 (the x-y plane), which leads to a non-deterministic state when the current is removed. It is also reported that the field-like torque $\tau_{FL}$ assists in the deterministic field-free SOT switching for single domain devices (macrospin mode) with a uniform magnetization [29]. In some cases, as shown in Figure 1(c), where the DMI alone is unable to achieve a deterministic field-free SOT switching, a small field-like torque $\tau_{FL}$ ($R = \dfrac{\tau_{FL}}{\tau_{SL}} = 0.05$) is able to assist in the successful magnetization switching.

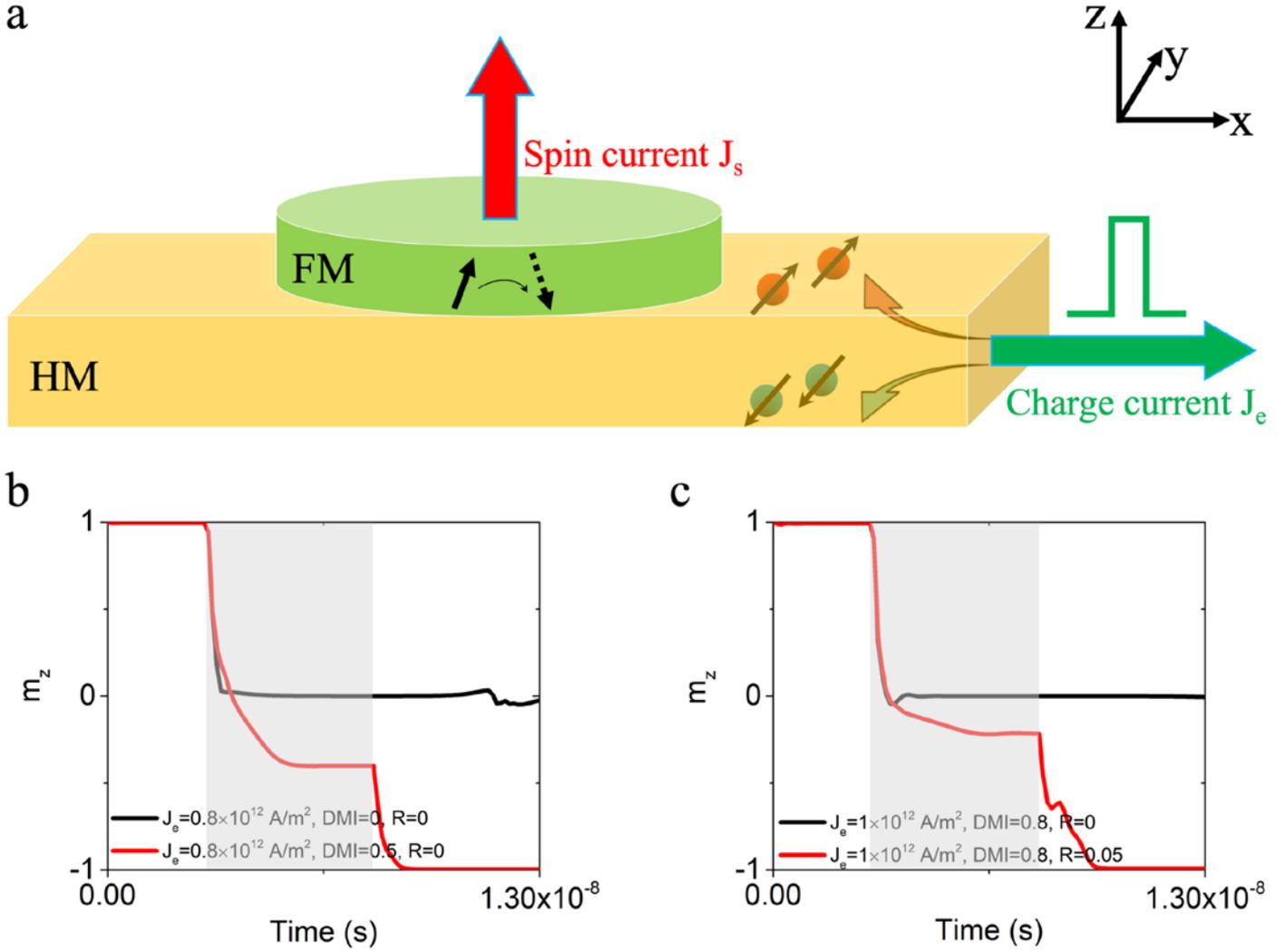

Figure 1. (a) A sketch of the field-free SOT switching in a HM/FM structure. The FM circular dot with perpendicular magnetic anisotropy is on top of a HM stripe. A FM circular dot with a diameter of 100 nm is assumed. Charge current pulses are injected along the HM channel (x-axis). (b) Temporal evolutions of the perpendicular magnetization component, $m_z$, under various DMI conditions ($DMI = 0$ and $0.5\ mJ/m^2$) without the assist of field-like torque $\tau_{FL}$ ($R = 0$). (c) Temporal evolutions of the perpendicular magnetization component, $m_z$, under various field-like torque conditions ($R = 0$ and $R = 0.05$) with DMI $= 0.8\ mJ/m^2$. The charge current pulse $J_e$ is applied during a time window of 5 ns (the shaded region).

3. Results

*3.1 Deterministic Field-free Switching via DMI*

In this section, we discuss the magnetization switching behaviors of the perpendicularly magnetized FM layer without considering the effect of field-like torque $\tau_{FL}$ (R=0). Figure 2 shows the temporal evolutions of out-of-plane magnetization component, $m_z$, under different combinations of charge current densities and DMI parameters. At low charge current density ($J_e = 0.5 \times 10^{12}\ A/m^2$), below the critical switching current, magnetization switching fails for both DMI and non-DMI cases. On the other hand, at high charge current density ($J_e = 1.5 \times 10^{12}\ A/m^2$), the magnetization is trapped in the x-y plane for both DMI and non-DMI cases. As is expected, deterministic field-free SOT switching of perpendicularly magnetized FM layer is unachievable without the DMI term ($\text{DMI} = 0\ mJ/m^2$). With intermediate charge current density and DMI terms, successful magnetization switching is achieved at different combinations of DMI and $J_e$ values.

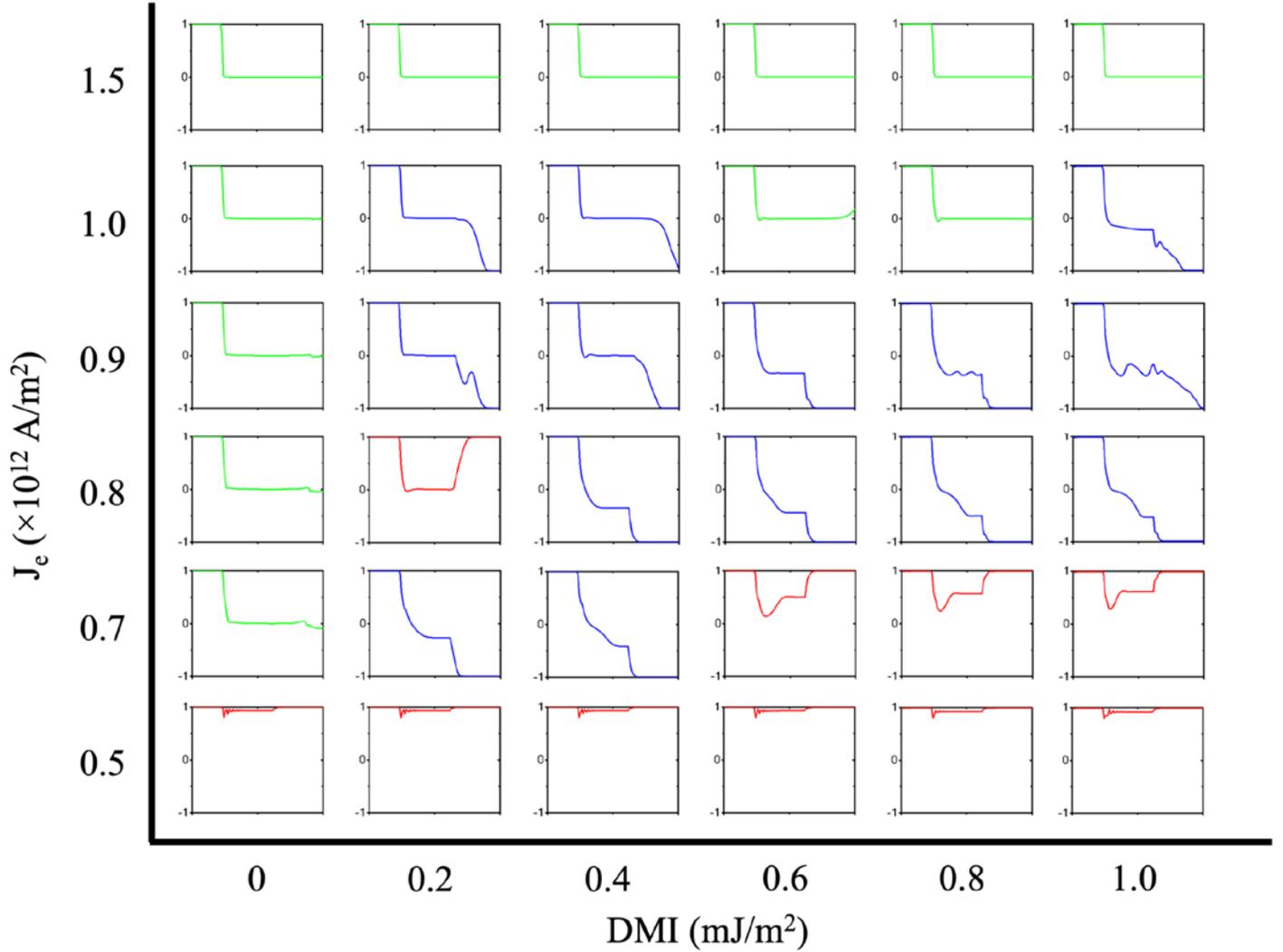

Figure 2. Field-free SOT switching of the perpendicularly magnetized FM circular dot without the effect of field-like torque $\tau_{FL}$. The temporal evolutions of the perpendicular magnetization component, $m_z$, under various DMI and $J_e$ conditions. Blue lines indicate successful magnetization switching, green lines indicate that the magnetization is switched to the x-y plane, and red lines indicate a fail magnetization switching where the magnetization returns to previous state (+z direction) after the current pulse.

We also perform simulations with other DMI and $J_e$ values to investigate the deterministic field-free switching without the field-like torque term $\tau_{FL}$. The charge current density $J_e$ varies from $0.5 \times 10^{12}$ to $1.5 \times 10^{12}$ A/m² with a step value of $0.1 \times 10^{12}$ A/m² and the DMI parameter varies from 0 to 1.0 mJ/m² with a step value of 0.1 mJ/m².

Figure 3(a) summarizes the deterministic field-free SOT switching results for different combinations of DMI and $J_e$ for a circular device of 100 nm diameter. Blue region is the deterministic field-free SOT switching window.

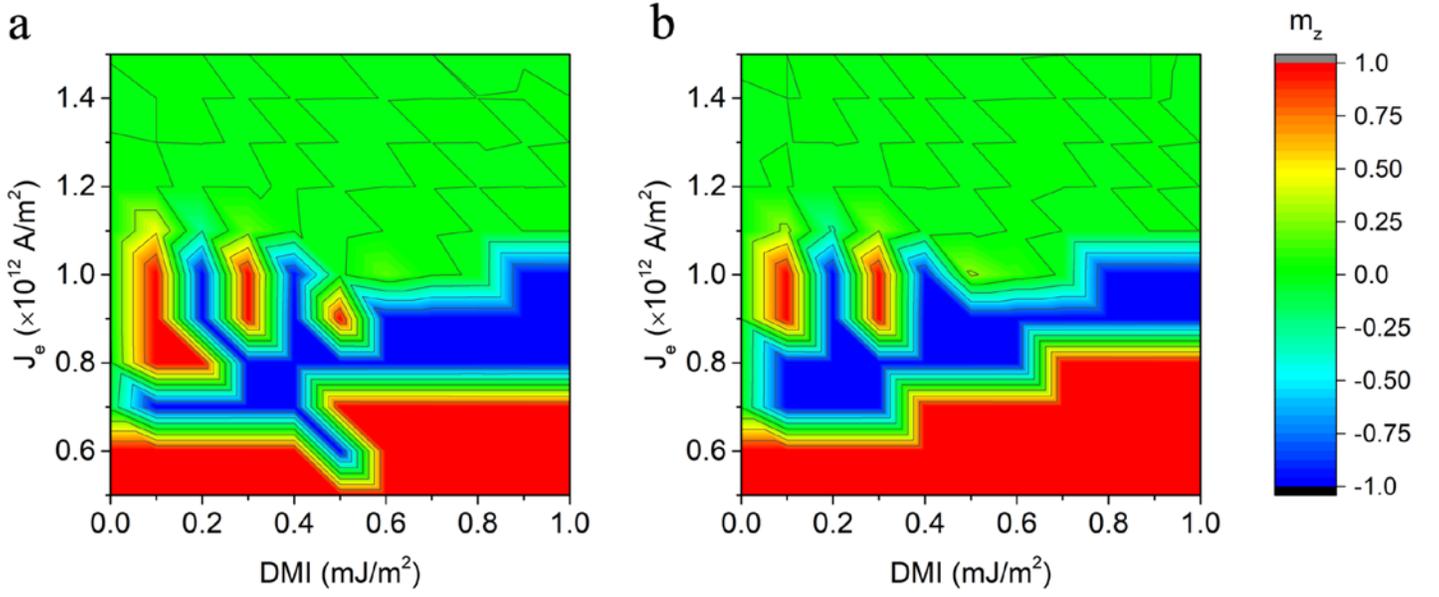

Figure 3. Deterministic field-free spin-orbit torque switching window for varying DMI values, charge current densities $J_e$, and field-like torque $\boldsymbol{\tau}_{FL}$ for the FM circular dot. (a) Without the field-like torque $\boldsymbol{\tau}_{FL}$. (b) With the field-like torque $\boldsymbol{\tau}_{FL}$, $R = \boldsymbol{\tau}_{FL}/\boldsymbol{\tau}_{SL} = 0.05$. Blue region indicates successful magnetization switching, green region indicates that the magnetization is switched to the x-y plane, and red region indicates a fail magnetization switching where the magnetization returns to previous state (+z direction) after the current pulse.

*3.2 Deterministic Field-free Switching via the Joint Effects of DMI and Field-like Torque*

In Figure 3(b), we added the field-like torque term $\boldsymbol{\tau}_{FL}$ with magnitude of $R = \boldsymbol{\tau}_{FL}/\boldsymbol{\tau}_{SL} = 0.05$ in the simulations and the deterministic field-free SOT switching results for different combinations of DMI and $J_e$ are summarized. By adding a small field-like torque term, the deterministic field-free SOT switching window expands compared with the results from Figure 3(a). Furthermore, we find that for a successful magnetization switching, as the DMI value increases, the critical charge current density $J_e$ increases. By adding a field-like torque term, the temporal evolutions of $m_z$ under different combinations of charge current densities and DMI parameters are plotted in Figure 4.

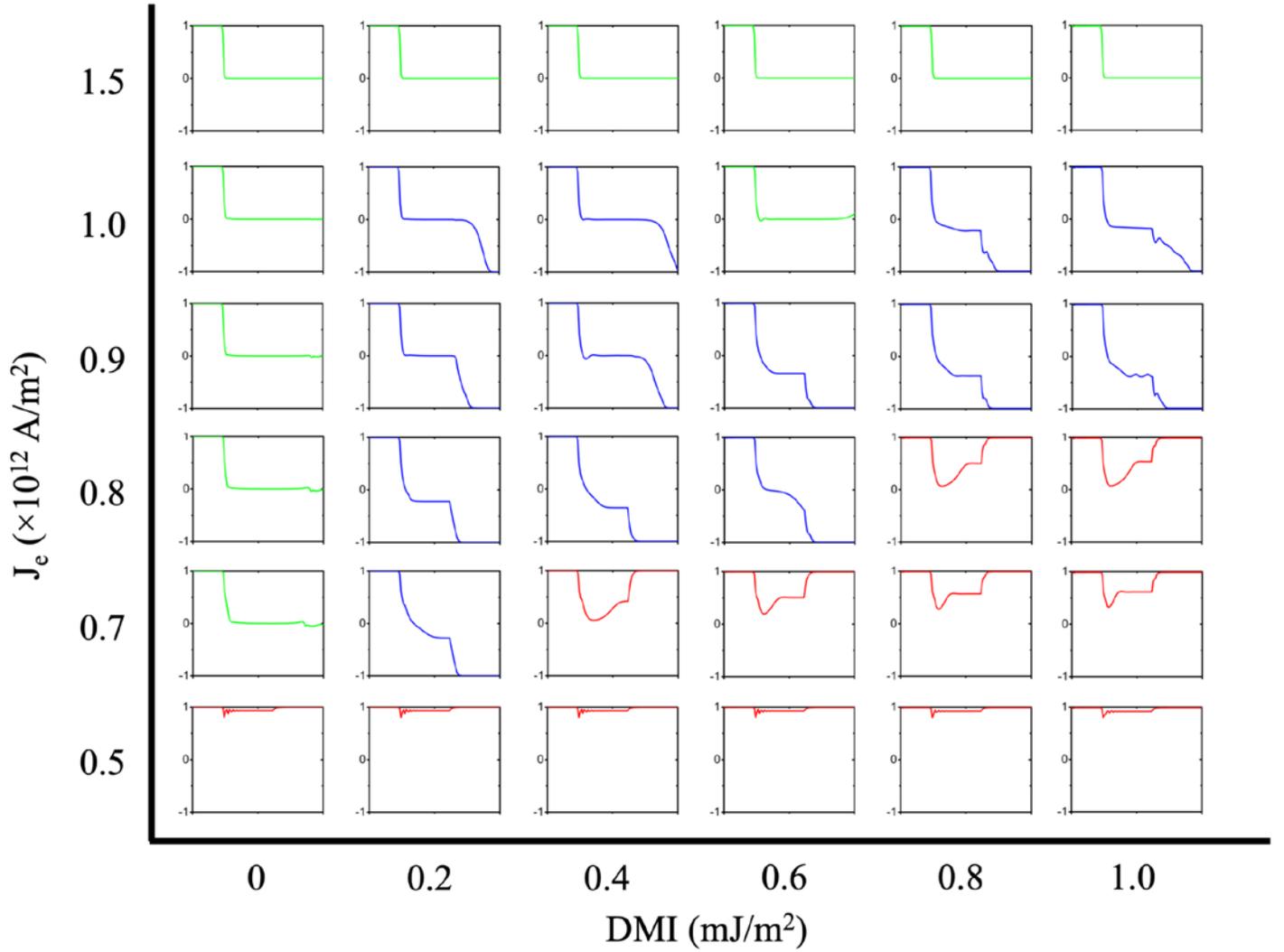

Figure 4. Field-free SOT switching of the FM circular dot in the presence of field-like torque $\tau_{FL}$, $R = \tau_{FL}/\tau_{SL} = 0.05$. The temporal evolutions of the perpendicular magnetization component, $m_z$, under various DMI and $J_e$ conditions. Blue lines indicate successful magnetization switching, green lines indicate that the magnetization is switched to the x-y plane, and red lines indicate a fail magnetization switching where the magnetization returns to previous state (+z direction) after the current pulse.

*3.3 The Effect of Field-like Torque on Deterministic Field-free Switching*

In this section, we explore the effect of field-like torque term $\tau_{FL}$ on the deterministic SOT field-free switching. The strength of field-like torque $\tau_{FL}$ is characterized by the ratio of $R = \tau_{FL}/\tau_{SL}$ and, herein, we increased the R from 0 to 0.3 for different combinations of DMI and $J_e$ values. Figure 5 shows the adverse effect of field-like torque $\tau_{FL}$ on the deterministic SOT field-free switching. When the charge current density and DMI are set as constants ($DMI = 0.5\ mJ/m^2$ and $J_e = 0.8 \times 10^{12}\ A/m^2$), successful magnetization switching is achieved without the assistance of field like torque ($R = 0$). Since DMI induces significant magnetization tilting at the edges of magnetic structures, resulting in asymmetric field-induced domain nucleation [30]. Figure 5(b) shows that the SOT-induced magnetization switching in the presence of DMI is governed by domain nucleation on one edge followed by propagation to the opposite edge. The switching time $t_0$, defined by $m_z|_{t_0} = 0$, increases as the $R$ increases, and the slope of $m_z(t)$ in Figure 5(a) indicates that this is related to a slower domain wall propagation. Which is also confirmed by the magnetization profiles in Figure 5(b) where the domain wall propagates at a slower rate in the second row ($R = 0.05$) compared to the third row ($R = 0$). As a result, as we increase the magnitude of field-like torque by increasing the $R$, the $m_z$ flips with a slower pace, and the magnetization fails to flip to -z direction as $R$ reaches to 0.1.

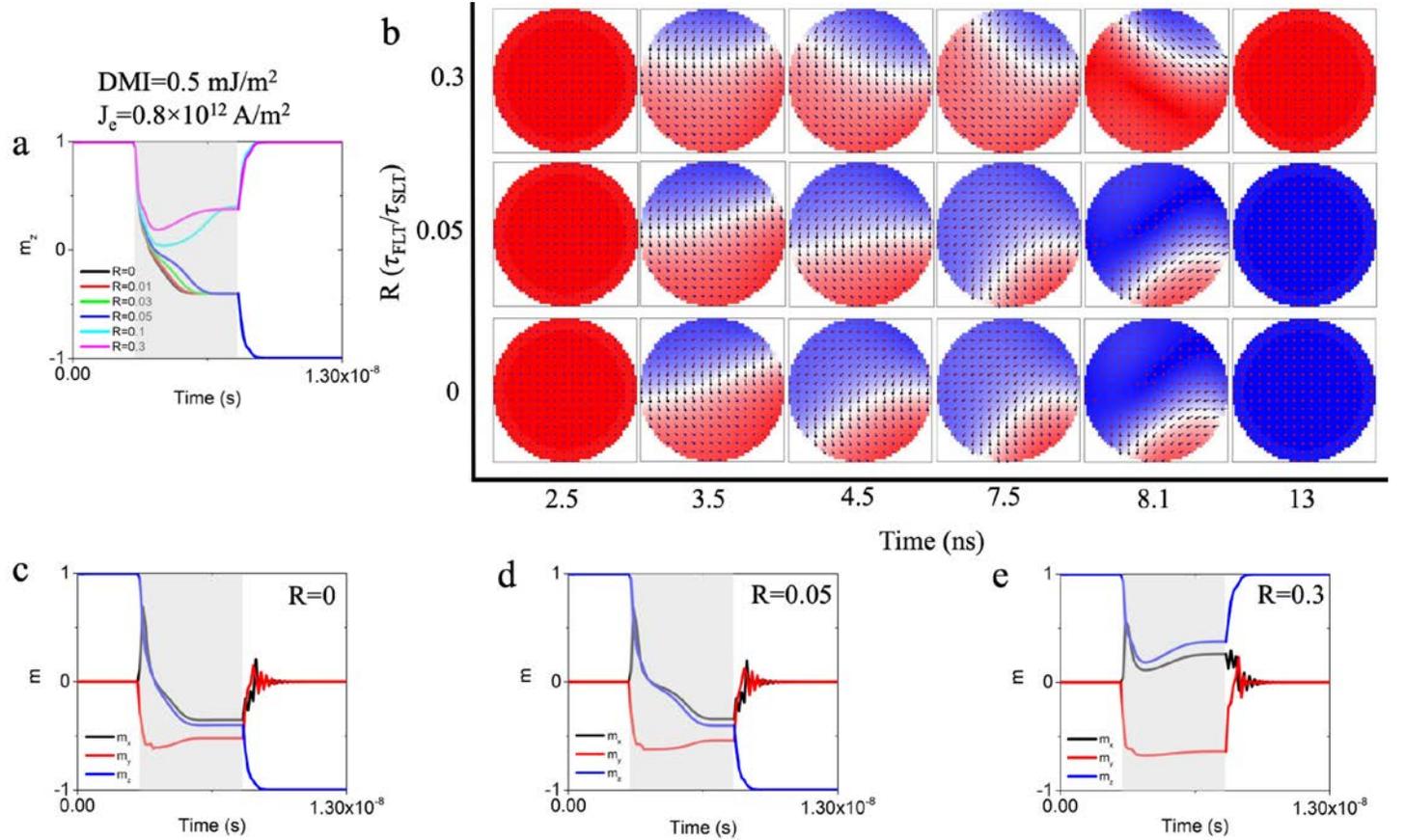

Figure 5. Adverse effect of the field-like torque on the field-free SOT switching in the FM circular dot in the presence of DMI (current pulse is applied within the shaded region). $DMI = 0.5\ mJ/m^2$ and $J_e = 0.8 \times 10^{12}\ A/m^2$. (a) The temporal evolutions of the perpendicular magnetization component, $m_z$, under various $R$ conditions. (b) The snapshots of magnetization profiles before (t=2.5 ns), during (t=3.5, 4.5, and 7.5 ns), and after (t=8.1 and 13 ns) the charge current pulse. (c) – (e) are the temporal evolutions of the $m_x, m_y$, and $m_z$ components during one magnetization switching period. (c) $DMI = 0.5\ mJ/m^2$, $R = 0$, and $J_e = 0.8 \times 10^{12}\ A/m^2$. (d) $DMI = 0.5\ mJ/m^2$, $R = 0.05$, and $J_e = 0.8 \times 10^{12}\ A/m^2$. (e) $DMI = 0.5\ mJ/m^2$, $R = 0.3$, and $J_e = 0.8 \times 10^{12}\ A/m^2$.

The adverse effect of field-like torque $\tau_{FL}$ on the deterministic SOT field-free switching is very commonly seen in this simulation. In addition, we find that the successful magnetization switching with the assistance of larger DMI values are more susceptible to the adverse effect of field-like torque $\tau_{FL}$. In Figure 6, we show another case where $DMI = 1.0 \ mJ/m^2$ and $J_e = 0.8 \times 10^{12} \ A/m^2$ can successfully switch the magnetization in FM layer without the assistance of field-like torque. Compared to the case in Figure 5, the DMI value is doubled and the charge current density is identical, however, the DMI-based field-free switching is blocked by increasing the magnitude of $\tau_{FL}$ to $R = 0.03$. Similarly, it is noted that the field-like torque slows down the speed of domain wall propagation from the magnetization profiles in Figure 6(b).

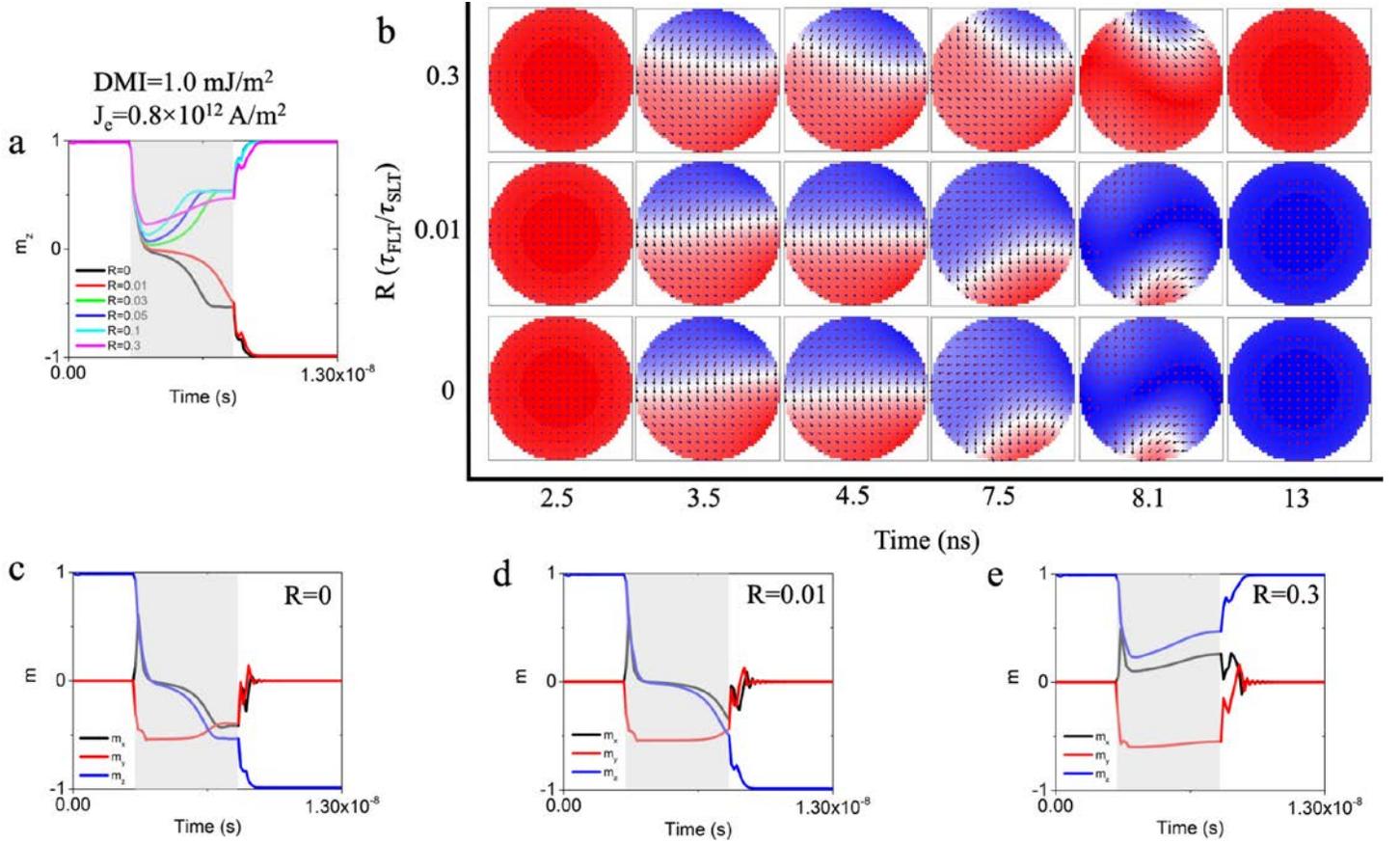

Figure 6. Adverse effect of the field-like torque on the field-free SOT switching in the presence of DMI (current pulse is applied within the shaded region). $DMI = 1.0 \ mJ/m^2$ and $J_e = 0.8 \times 10^{12} \ A/m^2$. (a) The temporal evolutions of the perpendicular magnetization component, $m_z$, under various $R$ conditions. (b) The snapshots of magnetization profiles before (t=2.5 ns), during (t=3.5, 4.5, and 7.5 ns), and after (t=8.1 and 13 ns) the charge current pulse. (c) – (e) are the temporal evolutions of the $m_x, m_y,$ and $m_z$ components during one magnetization switching period. (c) $DMI = 1.0 \ mJ/m^2$, $R = 0$, and $J_e = 0.8 \times 10^{12} \ A/m^2$. (d) $DMI = 1.0 \ mJ/m^2$, $R = 0.01$, and $J_e = 0.8 \times 10^{12} \ A/m^2$. (e) $DMI = 1.0 \ mJ/m^2$, $R = 0.3$, and $J_e = 0.8 \times 10^{12} \ A/m^2$.

On the other hand, in some rare cases, we found the supportive effect of field-like torque $\tau_{FL}$ on the deterministic SOT field-free switching. As shown in Figure 7, where the $DMI = 0.5\ mJ/m^2$ and $J_e = 0.9 \times 10^{12}\ A/m^2$ is unable to switch the magnetization without the assistance of field-like torque ($R = 0$). However, within a certain range of magnitudes, the field-like torque assists in the deterministic field-free switching ($R = 0.01, 0.03, 0.1$). While a larger field-like torque ($R = 0.3$) fails to reach deterministic switching.

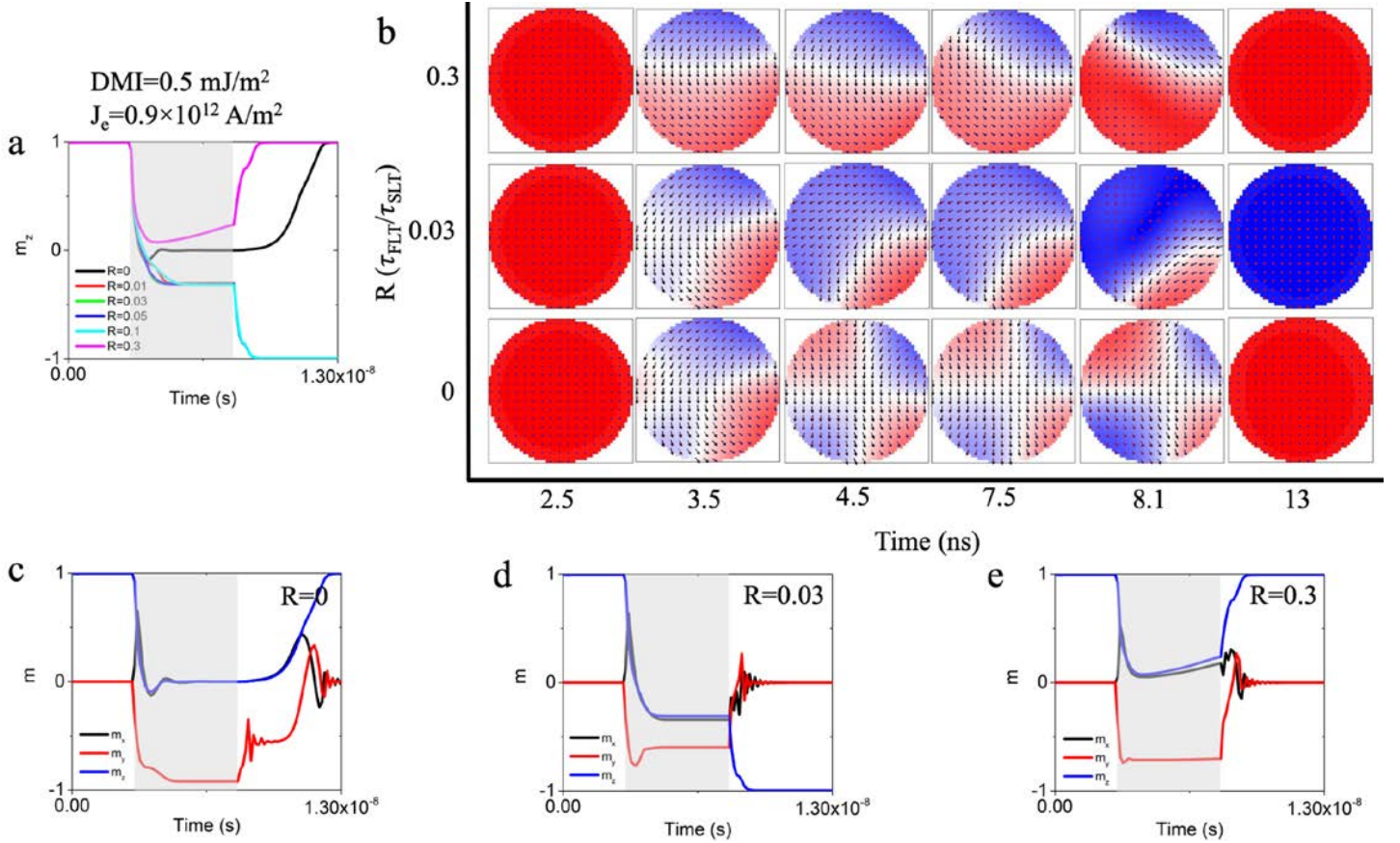

Figure 7. Supportive effect of the field-like torque on the field-free SOT switching in the presence of DMI (current pulse is applied within the shaded region). $DMI = 0.5\ mJ/m^2$ and $J_e = 0.9 \times 10^{12}\ A/m^2$. (a) The temporal evolutions of the perpendicular magnetization component, $m_z$, under various $R$ conditions. (b) The snapshots of magnetization profiles before (t=2.5 ns), during (t=3.5, 4.5, and 7.5 ns), and after (t=8.1 and 13 ns) the charge current pulse. (c) – (e) are the temporal evolutions of the $m_x, m_y$, and $m_z$ components during one magnetization switching period. (c) $DMI = 0.5\ mJ/m^2$, $R = 0$, and $J_e = 0.9 \times 10^{12}\ A/m^2$. (d) $DMI = 0.5\ mJ/m^2$, $R = 0.03$, and $J_e = 0.9 \times 10^{12}\ A/m^2$. (e) $DMI = 0.5\ mJ/m^2$, $R = 0.3$, and $J_e = 0.9 \times 10^{12}\ A/m^2$.

4. **Conclusions**

In this work, we have performed micromagnetic simulations on a FM/HM structure to investigate the magnetization dynamics for a deterministic field-free SOT switching in perpendicularly magnetized FM circular dot. Different switching processes and results are identified for different charge current densities, DMI values, and FLT values. The effects of DMI and current density on deterministic field-free switching performance are presented. We have mapped the current density window for deterministic switching via DMI only and via the joint effects of DMI and FLT for a FM circular dot of 100 nm diameter. It is found that the FLT can effectively expand the current density window for deterministic field-free SOT switching. We have also shown that the FLT can play an adverse role in blocking and slowing down the deterministic field-free switching via DMI. However, on the other hand, FLT is able to assist DMI in achieving a successful deterministic field-free switching.

**Acknowledgements**

This study was financially supported by the Institute of Engineering in Medicine of the University of Minnesota, National Science Foundation MRSEC facility program, the Distinguished McKnight University Professorship, Centennial Chair Professorship, Robert F Hartmann Endowed Chair, and UROP program from the University of Minnesota.

**Conflict of Interest**

The authors declare no conflict of interest.